# Kolmogorov complexity of sequences of random numbers generated in Bell's experiments.

Marcelo G. Kovalsky, Alejandro A. Hnilo and Mónica B. Agüero
*CEILAP, Centro de Investigaciones en Láseres y Aplicaciones, UNIDEF (MINDEF-CONICET);*
*CITEDEF, J.B. de La Salle 4397, (1603) Villa Martelli, Argentina.*
email: ahnilo@citedef.gob.ar

May 18th, 2018.

Quantum systems are the ultimate touchstone for the production of random sequences of numbers. Spatially spread entangled systems allow the generation of identical random sequences in remote locations. The impossibility of observing a quantum system, without disturbing it, ensures that the messages encoded using these sequences cannot be eavesdropped. This is the basis of Quantum Key Distribution. It is then of crucial importance knowing whether the sequences generated in the practice by spatially spread entangled states are truly random, or not. Yet, that knowledge is not immediate. One of the obstacles is the very definition of randomness. "Statistical" randomness is related with the frequency of occurrence of strings of data. On the other hand, "algorithmic" randomness is related with the compressibility of the sequence, what is given by Kolmogorov complexity. We analyze sequences generated by entangled pairs of photons focusing on an estimation of their complexity.

Keywords: QKD and communication security. Kolmogorov complexity. Randomness of time series.

## 1. Introduction.

Sequences of random numbers are a basic supply in many applied sciences of information, from statistics to cryptography. The randomness of a given sequence is difficult to establish in the practice. Even the very definition of "random" is controversial. All definitions agree that "predictable" ⇒ "not random", hence "random" ⇒ "unpredictable". But, the unpredictability of an event is, in general, an ambiguous property. It depends on the available information. Some consensus has been reached, that appropriate measurements performed on quantum systems guarantee randomness. This is a consequence of von Neumann's axiom: quantum measurements violate the principle of sufficient reason. Or, in other words, a quantum measurement produces one or another outcome *without cause*. It is intuitive to conclude that a sequence of such outcomes is unpredictable, although this conclusion is difficult to formalize [1]. Besides, note that it is still logically possible the existence of sequences that are both "unpredictable" and "not random". Depending on the precise definitions of "random" and "predictable" involved, chaotic series may be an example of this case.

Leaving aside the logical intricacies, there are at least two definitions of randomness that are relevant from a practical point of view [2]:

*i)* "Statistical" randomness. Imagine a sequence of 1 and 0. The sequence is "statistically random" if the number of strings of 1 and 0 of different length *n* (say, 110101 for *n*=6), in the total sequence, coincides with the number one would get if the sequence had been produced by tossing an ideal coin (statistical spread is taken into account, of course). Yet, certifying this property for any value of *n* and/or different ways of choosing the strings is not easy. Other tests of statistical randomness involve the decay of the self-correlation or the mutual information. They all involve measuring probabilities. The battery of tests provided by the National Institute of Standards and Technology (NIST) is mostly based on this approach.

*ii)* "Algorithmic" randomness. A sequence can pass the tests mentioned above and still be predictable, hence, not random. A well known example is the sequence formed by the digits of the number π (or any other transcendental number). A sequence is "algorithmically random" if there is no algorithm or program code able to generate the sequence using a number of bits shorter than the said sequence. Note that this definition does not use probabilities. It applies even to sequences that are not statistically stationary. By the way, practical tests of randomness often include subroutines aimed to recognize the digits of the best known transcendental numbers.

Algorithmic randomness is directly related with the definition of *complexity* developed by Kolmogorov [3], Chaitin [4] and Solomonoff [5]. In few words, the complexity *K* of a binary sequence of length *N* is the binary length of the shortest program (running on a classical Turing machine) whose output is the said binary sequence. A sequence is "algorithmically random" if $K \approx N$. As there is no way of expressing the sequence using less bits that the sequence itself, the sequence is said to be *incompressible*. This definition is intuitive and appealing, but it has two main drawbacks. One: it is possible to demonstrate that all series are (partially) compressible; hence, that the precise condition $K \approx N$ cannot be reached. In the practice this problem is solved by appropriately rescaling the definition. Two: *K* cannot be actually *computed*, for one can never be sure that there is no shorter program able to generate the sequence. Nevertheless, *K* can be *estimated* from the compressibility of the sequence using, f.ex., the algorithm devised by Lempel and Ziv [6].

Measurements performed on quantum entangled states can generate identical outcomes in two remote stations. The sequences of outcomes, assumed random, allow the encryption of messages in a secure way using one-time-pad Vernam's cipher. This technique is known as Quantum Cryptography [7], or Quantum Key Distribution (QKD). The purity of the achieved entanglement puts a minimum bound on the entropy of the generated sequences [8] and hence, to the degree of statistical randomness. The loophole-free verification of the violation of Bell's inequalities was required as a necessary step to certify the randomness of the sequences and the invulnerability of QKD [1]. This loophole-free verification has been recently achieved by several groups using different techniques [9-12] (for a sort of critical review, see [13]). Loophole-free generated sequences have been recently used to produce series of numbers of "quantum certified" statistical randomness [14]. To our knowledge, little or no attention has been paid to algorithmic randomness.

In this paper, we study the algorithmic randomness of time series generated in Bell's experiments by using the realization of the Lempel and Ziv algorithm developed by Kaspar and Schuster [15]. We also use part of the battery of tests of NIST to evaluate the statistical randomness of the same files. It is evident that the results of these tests, performed on *actual* Bell's sequences, are crucial to ensure the invulnerability of QKD in the practice. In the Section 2, we briefly describe the idea in Lempel and Ziv algorithm. We also review some previous attempts to detect deviations from randomness in Bell's series. In the Section 3 we report the results for the main set of data of the Bell's experiment performed in Innsbruck in 1998 [16], generously provided by Prof. G.Weihs. We also include some data recorded in our own setup. Although our experiment is far more modest, it puts some light on the probable cause of the regularities found in some runs of the Innsbruck experiment.

## 2. Background.
*2.1 Lempel and Ziv algorithm.*

Complexity has advantages over other methods of detecting regular behavior in a time series. Regarding the statistical methods, complexity does not need to assume stationary probabilities. Regarding non-statistical methods, as the ones extracted from the theory of nonlinear dynamics systems (Takens' theorem), complexity does not need to assume the existence of a low dimensional object in phase space. On the other hand, complexity cannot be properly calculated; it can only be estimated.

Assuming a time series of elements $\{s_1, s_2 \ldots s_N\}$ the Lempel and Ziv algorithm adds a new "word" to its memory every time it finds a substring of consecutive elements not previously registered. The size of the compiled vocabulary, and the rate at which new words are found, are the basic ingredients to evaluate complexity. In the realization of the algorithm [15, 17], the time series is encoded so that a binary string is produced. Then the complexity counter $c(N)$, which is defined as the minimum number of distinct words in a given sequence of length $N$, is calculated. As $N \to \infty$, $c(N) \to N/log_2(N)$ in a random series. The normalized complexity measure $K$ is then defined as:

$$K(N) \equiv c(N) \times log_2(N)/N \qquad (1)$$

The value of $K(N)$ is designed to be near to 0 for a periodic or regular time series, and 1 for a random one, if the available value of $N$ is large enough. For a chaotic series it is typically between 0 and 1. For a "strongly" random sequence of relatively short length, $K(N)$ can be considerably larger than 1. As references, the series formed by the digits of the number $\pi$ has $K(27,000) = 0.95$. A typical chaotic time series (dimension of embedding $d_E = 4$, one positive Lyapunov exponent) recorded from a solid-state laser with modulated losses [18] has $K(10^5) = 0.4$. A numerically generated quasi-periodical (2-torus) sequence has $K(10^6) < 0.02$.

*2.2 Previous studies of randomness in Bell's series.*

Some years ago, we looked for regularities in the time series generated in the Innsbruck experiment. This experiment is not only crucial to the foundations of Quantum Mechanics, but also is a superb realization of the quantum channel of a QKD setup.

In that experiment, each *run* includes 4 *files*, that is: for each of the two stations, one has the time of photon detection, and also a code for the angle setting and fired detector (see Fig.1). We firstly looked for periodicities in one of the runs (named *longdist35*) using standard linear transforms [19], finding none. Later, we sought for low dimensional objects in phase space, using Takens' reconstruction theorem [20], on the whole set of available data [21]. We found a chaotic attractor with $d_E=10$, and four positive Lyapunov exponents, in the longest run in real time. It is named here *longtime*, and is made of the runs originally named *longtime1* and *newlongtime2*. It was possible to reconstruct the attractor and to predict the outcomes in the sequence roughly up to the inverse of the largest positive Lyapunov exponent, as expected. Remarkably, the same was possible for the 16 subsets corresponding to the different settings in spite of their shorter length. If the files in *longtime* were used for QKD, it would be possible to predict until 20 bits of the key, what is a vulnerability of a new kind [21]. The run *longtime* was the only one where $d_E$ was reliably measured. In order to check if the cause was its time length, we perform a simpler Bell's experiment, but with an unusually long continuous time of observation. It amounts to more than half an hour, about five times longer than *longtime*. In this run, named here *SL1722*, no value of $d_E$ is reliably measured. The cause of the attractor in *longtime* is believed to be a drift between Alice and Bob's clocks. File *SL1722* is recorded with a single clock instead, so that obtained results are consistent with this belief. Unfortunately, when the attractor in *longtime* was found the Innsbruck experiment had been dismantled, so it was impossible to know its cause by sure.

## 3. Results.
*3.1 Structure of the Innsbruck experiment's runs.*

The Innsbruck experiment includes fast switching of the analyzers' settings, driven by independent and quantum-measurement-based random number generators, and spatially distant stations, what is named the "remote, switched" condition. Most of the results obtained in this condition are the set of runs named *longdist**. We also study some preparatory runs with the stations close to each other and slowly varying settings (condition "local, static"). Also, with close stations and fast and random switched settings (condition "local, switched"). There are no "remote, static" runs. We discard most of the runs that do not violate the involved Bell's inequality ($S_{CHSH} \leq 2$).

```
2.1634050886170270e-006   1
8.0075823256314910e-006   1
1.2668053500635000e-005   2
4.9706368996378400e-005   2
5.2854269101605610e-005   0
9.7272343207776580e-005   2
1.2815431751139350e-004   3
1.3008972522198680e-004   0
1.4427547709393630e-004   0
1.5615472722963800e-004   3
2.0560825198241920e-004   0
2.1648761420145820e-004   3
2.1938141279290700e-004   2
2.8082761420145820e-004   3
2.9832825198241920e-004   0
3.1709738886421220e-004   0
3.2956761420145820e-004   3
3.3093472722963800e-004   2
3.4241627025788890e-004   0
3.5539167101916270e-004   1
3.6150337933573870e-004   2
3.6763245357770900e-004   2
3.8984212241744120e-004   1
4.3617738345535160e-004   3
4.4365097917360800e-004   2
4.5388049708441920e-004   1
4.9135137159675660e-004   1
4.9907703051256600e-004   2
5.2821928900505510e-004   1
```

Figure 1: Files in the Innsbruck experiment. The left column (*_V.dat file) indicates the time photons were detected, in seconds. The right column (*_C.dat file) indicates the detector that fired and the analyzer's orientation, according to a code. The data belong to *Alice* station of run *longdist35*.

The structure of the runs is shown in the Figure 1. For each of the two stations, there are two files: the one with extension *_V.dat (left column in the Figure) is the (always increasing) sequence of photon detection times, in seconds. The one with the same name but extension *_C.dat (right column in the Figure) indicates the setting of the analyzer and the detector that fired at that time, using a two-bit code. Both files have the same length. There is a pair of similar files for the other station. The length of the files is the number of single photons detected, and is hence different in each station.

A coincidence occurs when the difference of the values in the *_V.dat for each station is smaller than a certain value, what is called "time coincidence window" $T_w$. Once a coincidence is found, we pick up the time value in *_V.dat of station *Alice* (this choosing is arbitrary, it may well to be *Bob*, or the average between them, in any case the difference is small) to write down a time sequence of coincidences. The corresponding codes in the two C.dat files allow calculating the value of the $S_{CHSH}$ parameter.

In time stamped setups like these, the value of $T_w$ can be chosen at will after the experiment has ended. Due to different response times of the detectors and electronic channels, cable lengths, etc. a time delay between the files in each station must be added. The value of the delay is found by maximizing the number of coincidences for a given value of $T_w$. This leads to some ambiguity in the definition of the coincidences' file. Here we use the values of $T_w$ and delay reported by the authors of the experiment.

*3.2 Algorithmic and statistic randomness.*

We calculate *K* (Kaspar and Schuster realization of Lempel and Zev algorithm) and also submit the runs to a battery of randomness tests developed by the NIST. The complete battery includes 15 tests. Here we use the simplest 6, namely: Frequency (Monobit) Test, Frequency Test within a Block, Runs Test, Tests for the Longest-Run-of-Ones in a Block, Binary Matrix Rank Test and Discrete Fourier Transform (Spectral) Test. A run is said to have positive ("yes") randomness only if passes the 6 tests. The calculation of *K* and these 6 tests form a set of relatively simple and fast running programs that are feasible to be included as a control in a QKD setup in the practice. As will be seen, all runs that are discarded by the criterion *K*<1 are also discarded by (at least one of) the NIST tests. This result is specific of the set of runs included in this study. Given the different nature of the two types of randomness, the safe criterion is that a sequence can be considered random only if it passes *both* types of tests.

The main results are summarized in the Table 1. The last column is the length of the sequence. It corresponds to the number of coincidences, excepting for the "singles" files, in which case they correspond to the *Alice* station. There are three groups of sequences with different complexities (Fig.2): the ones with *K*<1, which cannot be considered random, the ones with *K*≈1, which are "normally" random, and the ones with *K*>1, which are "strongly" random, what means that the normalization factor in eq.(1) is insufficient. Most of the sequences belong to the latter two groups, meaning that Bell's experiments do generate series of algorithmically random numbers, as expected.

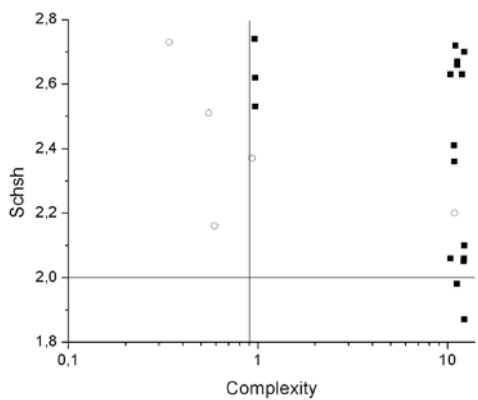

Figure 2: Graphical representation of data in Table 1. Open circles indicate the runs that do not pass NIST tests. Horizontal line indicates the Bell's limit, vertical line *K*=0.9.

The run *longtime* has not only low *K* and is discarded by the NIST statistical tests (3 over 6), but is even partially predictable, as it was discussed before. The subset corresponding to analyzers' settings and firing detectors *Alice*=0, *Bob*=3 shows a slightly higher *K* than the complete sequence (probably because it is shorter), and is also unable to pass NIST. The runs

*longdist22* and *longdist35* have low *K* and, correspondingly, they do not pass NIST. Runs *longdist10* and *12* have high *K* but do not pass NIST. None of these five runs can be considered random despite they violate the involved Bell's inequality (they are the open circles in Fig.2)

All the runs with *K*>1 pass the NIST excepting *longdist10*. As a reference, run *Conlt3* is obtained from the coincidences between detectors observing uncorrelated fields, it has *K*=7.29 and passes NIST.

It seems that higher *K* corresponds to lower $S_{CHSH}$. F.ex., the three runs with higher *K* (*longdist30, 32* and *37*, all with *K*>12) have an average $S_{CHSH}$=2.28, while the three ones with lower *K* (but still *K*≈1, *longdist0, 36* and *31*) have an average $S_{CHSH}$=2.62. Runs *longdist10* and *12* are not included in this set because they do not pass NIST. Run *longdist34*, which has the highest complexity of all (12.26), is also discarded because it has $S_{CHSH}$= 1.87 < 2.

The complexity of the singles files and the coincidence files is, in general, nearly the same. There are exceptions: in runs *longdist22* and *35* the value of *K* in the coincidence files is smaller than in the singles ones, down to the point that they cannot be considered random. On the other hand, in *longdist36* the value of *K* is larger for the coincidences than for the singles, although both can be considered random.

All runs obtained in the "local" condition (regardless if "switched" or "static") have *K*≈1 and pass the NIST. Finally, the sequences of settings and outputs (the column on the right in Fig.1) have high *K* and also pass NIST. This confirms the reliability of the random number generators used to drive the settings in the Innsbruck experiment.

**Summary.**

An estimation of Kolmogorov complexity in time series recorded in Bell's experiments has been performed. Almost all series have complexity *K*≈1 or *K*>1, what means they are algorithmically random. The few with *K*<1 belong to the "remote, switched" case. They do not pass NIST tests either. The deviation from the expected randomness is presumably caused by a drift between the clocks, an artifact that had been independently detected.

Even though low *K* does not allow, by itself, to predict outcomes, it implies that the involved sequences are compressible, and hence potentially vulnerable. In our opinion, the main conclusion of this study is that, although random sequences are generated in most cases, it is not safe taking randomness *for granted* in experimentally generated sequences, even if they violate the related Bell's inequality by a wide margin with a maximally entangled state. Deviations from randomness are observed even in the controlled conditions of the Innsbruck experiment, which are very difficult (perhaps impossible) to achieve in a QKD setup operating in a real world situation. Therefore, applying additional statistical and algorithmic tests and, if necessary, using distillation and extraction techniques are advisable before coding a message in the practice.

**Acknowledgments.**

Many thanks to Prof. Andrei Khrennikov for the encouragement, his observations and advices. Also, to Prof. Dragutin Mihailovic for his help to use the algorithms and to interpret their outputs. Many thanks again to Prof. Gregor Weihs for providing the files of the Innsbruck experiment. This work received support from the grants N62909-18-1-2021 Office of Naval Research Global (USA), and PIP 2017 0100027C CONICET (Argentina).

**References.**

[1] A.Khrennikov, "Randomness: Quantum vs classical", *Int.J.of Quantum Inform.* **14**, 1640009 (2016).
[2] G.Sommazzi, "Kolmogorov Randomness, Complexity and the Laws of Nature", https://www.researchgate.net/ publication /311486382.
[3] A.Kolmogorov, "Three approaches to the quantitative definition of information." *Problems of Information Transmission*, **1** p.4 (1965).
[4] G.Chaitin, "On the length of programs for computing binary sequences", *J. of the Association for Computing Machinery*, **13** p.547 (1966).
[5] R.Solomonoff, "A formal theory of inductive inference. part I", *Information and Control*, **7** p.1 (1964).
[6] A.Lempel and J.Ziv, *IEEE Trans. Inform. Theory* **22**, p.75 (1976).
[7] A.Ekert, "Quantum cryptography based on Bell's theorem"; *Phys.Rev.Lett* **67**, p.661 (1991).
[8] S.Pironio *et al.* "Random numbers certified by Bell's theorem", *Nature* **464** p.1021 (2010).
[9] M.Giustina *et al.*, "A Significant Loophole-Free Test of Bell's Theorem with Entangled Photons", *Phys. Rev. Lett.* **115**, 250401 (2015).
[10] L.Shalm *et al.*, "A Strong Loophole-Free Test of Local Realism", *Phys. Rev. Lett.* **115**, 250402 (2015).
[11] B.Hensen *et al.*, "Loophole-free Bell inequality violation using electron spins separated by 1.3 kilometers", *Nature* **526**, 682 (2015).
[12] W.Rosenfeld *et al.*, "Event-ready Bell-test using entangled atoms simultaneously closing detection and locality loopholes", *Phys.Rev.Lett.* **119**, 010402 (2017).
[13] A.Hnilo, "Consequences of recent loophole-free experiments on a relaxation of measurement independence"; *Phys. Rev. A* **95**, 022102 (2017).
[14] P.Bierhorst *et al.*, "Experimentally generated randomness certified by the impossibility of superluminal signals", *Nature* **556**, p.223 (2018).
[15] F.Kaspar and H.Schuster, "Easily calculable measure for the complexity of spatiotemporal patterns", *Phys.Rev.A* **36** p.842 (1987).
[16] G.Weihs *et al.*, "Violation of Bell's inequality under strict Einstein locality conditions", *Phys. Rev. Lett.* **81** p.5039 (1998).
[17] D.Mihailovic *et al.*, "Novel measures based on the Kolmogorov complexity for use in complex system behavior studies and time series analysis", *Open Phys.* 2015 13:1-14.
[18] N.Granese *et al.*, "Extreme events and crises observed in an all-solid-state laser with modulation of losses"; *Opt.Lett.* **41** p.3010 (2016).

[19] A.Hnilo et al., "Local realistic models tested by the EPRB experiment with random variable analyzers", Found. Phys.Lett. **15** p.359 (2002).
[20] H.Abarbanel, "Analysis of observed chaotic data", Springer-Verlag, New York, 1996.
[21] A.Hnilo et al., "Low dimension dynamics in the EPRB experiment with random variable analyzers", Found.Phys. **37** p.80 (2007).

| **Filename** (description) | Complexity | NIST (RND=?) | $S_{CHSH}$ | $N$ |
|---|---|---|---|---|
| **Longtime** (remote, switched) | 0.55 | NO | 2.51 | 95801 |
| **Longtime**, subset {0,3} | 0.65 | NO | Not applicable | 2122 |
| **Longdist0** (remote, switched) | 0.97 | yes | 2.53 | 15501 |
| **Longdist0**, singles | 0.96 | NO | Not applicable | 471017 |
| **Longdist1** | 11.94 | yes | 2.63 | 16168 |
| **Longdist2** | 11.21 | yes | 1.98 | 26675 |
| **Longdist3** | 11.25 | yes | 2.67 | 24335 |
| **Longdist4** | 11.24 | yes | 2.66 | 25402 |
| **Longdist10** | 10.88 | NO | 2.20 | 26529 |
| **Longdist11** | 10.82 | yes | 2.41 | 25573 |
| **Longdist12** | 0.93 | NO | 2.37 | 27158 |
| **Longdist12**, singles | 0.97 | yes | Not applicable | 934979 |
| **Longdist13** | 10.84 | yes | 2.36 | 27160 |
| **Longdist20** | 10.37 | yes | 2.06 | 41549 |
| **Longdist22** | 0.59 | NO | 2.16 | 39915 |
| **Longdist22**, singles | 0.96 | yes | Not applicable | 1237058 |
| **Longdist23** | 10.37 | yes | 2.63 | 41058 |
| **Longdist30** | 12.24 | yes | 2.10 | 14145 |
| **Longdist31** | 0.97 | yes | 2.62 | 13022 |
| **Longdist32** | 12.24 | yes | 2.70 | 10992 |
| **Longdist33** | 12.18 | yes | 2.06 | 13004 |
| **Longdist34** | 12.26 | yes | 1.87 | 14289 |
| **Longdist35** | 0.34 | NO | 2.73 | 14562 |
| **Longdist35**, singles | 0.96 | yes | Not applicable | 388455 |
| **Longdist36** | 11.0 | yes | 2.72 | 14573 |
| **Longdist36**, singles | 0.96 | yes | Not applicable | 388573 |
| **Longdist37** | 12.16 | yes | 2.05 | 14661 |
| **Loccorr1** (local, switched) | 0.96 | yes | 2.74 | 72533 |
| **Loccorr3** | 0.96 | yes | 2.74 | 73269 |
| **Loccorr3**, singles | 0.96 | yes | Not applicable | 853985 |
| **Bluesin1** (local, static), $\alpha=0°$, $\beta=7.5°$ | 0.98 | yes | Not applicable | 6797 |
| **Bluesin2**, $\alpha=0°$, $\beta=15°$ | 0.97 | yes | Not applicable | 6815 |
| **Bluesin3**, $\alpha=0°$, $\beta=22.5°$ | 0.97 | yes | Not applicable | 6822 |
| **Bluesin4**, $\alpha=0°$, $\beta=30°$ | 0.96 | yes | Not applicable | 6824 |
| **Bluesin5**, $\alpha=0°$, $\beta=37.5°$ | 0.97 | yes | Not applicable | 6784 |
| **SL1722** (local, static) $\alpha=0°$, $\beta=22.5°$ | 0.96 | yes | Not applicable | 56913 |
| **Conlt3** (local, static, uncorrelated) | 7.29 | yes | Not applicable | 4950 |

TABLE 1: Summary of results. They correspond to total coincidences between stations, unless indicated otherwise. The condition of the experiment is indicated for the first run with the same name, f..ex.: the condition of being "remote, switched" applies to all runs whose names start with *Longdist*. The "static" runs have fixed settings, which are indicated. All runs belong to the Innsbruck experiment, excepting *SL1722* and *Conlt3*, which are ours. The second column is "yes" only if the run passes all the 6 test of NIST named in the main text.